\documentclass[prb,aps,twocolumn,showpacs,floats]{revtex4}
\usepackage{psfig}

\begin{document}

\title{Doping change and distortion effect on
 double-exchange ferromagnetism}

\author{Phan Van-Nham and Tran Minh-Tien}
\affiliation{
Institute of Physics and Electronics, VAST,
P.O. Box 429, Boho, 10000 Hanoi, Vietnam.
}


\begin{abstract}
Doping change and distortion effect on the double-exchange
ferromagnetism are studied within a simplified double-exchange
model. The presence of distortion is  modelled by introducing the
Falicov-Kimball interaction between itinerant electrons and classical
variables. By employing the dynamical mean-field theory the charge
and spin susceptibility are exactly calculated. It is found that
there is a competition between the double-exchange induced
ferromagnetism and disorder-order transition. At low temperature
various long-range order phases such as charge
ordered and segregated phases coexist with ferromagnetism
depending on doping and distortion. A rich phase diagram is obtained.

\end{abstract}



\maketitle

\section{Introduction}

The discovery of colossal magnetoresistance in doped
manganites\cite{Salamon,Dagotto} has renewed interest in the
ferromagnetism induced by the double-exchange (DE) mechanism.\cite{Zener}
The main feature of the DE mechanism is a cooperative effect where electron
hoping favors ferromagnetic (FM) ordering of localized spins via
the FM Hund coupling, and vice versa, the presence of the FM order
facilitates the electron hoping. The occurrence of metallic FM state
in doped manganites R$_{1-x}$A$_{x}$MnO$_{3}$ (where R is  a
trivalent rare-earth element and A is a divalent alkaline ion) was
qualitatively explained by the DE mechanism.\cite{Furukawa} The
physically relevant electrons in these compounds are those from
Mn $3d$ levels  which are split
by the cubic crystalline field into triply degenerate $t_{2g}$
levels and doubly degenerate $e_{g}$ levels. Electrons of $e_{g}$
levels are able to hop between Mn sites and form the
conduction band, while electrons of $t_{2g}$ levels are localized.
Conduction electrons and localized spins are correlated by the DE
mechanism which leads to the appearance of the metallic FM phase. The
DE model became the starting point toward comprehensive
understanding of the properties of doped manganites.

Experiments have observed in doped manganites a very rich phase
diagram, which involves phases with spin, charge and orbital
orders.\cite{Salamon} For the undoped case ($x=0$) all Mn ions are
Mn$^{3+}$ and are expected to induce a Jahn-Teller (JT)
distortion. For the other extreme doping ($x=1$) all Mn ions are
Mn$^{4+}$ and do not couple to the JT distortion.  In the regime of
intermediate doping, two valence ions Mn$^{3+}$ and Mn$^{4+}$ are
simultaneously present. The presence of two valence ions may
lead to a static mixed valence Mn$^{3+}$/Mn$^{4+}$ configuration,
in particular, to an alternation charge-ordered (CO) state of
Mn$^{3+}$/Mn$^{4+}$ ions for appropriate dopings. This is the
conventional view, for which there are abundant experimental and
theoretical supports.\cite{Salamon,Dagotto}  In particular,
recently a CO-FM state has been observed.\cite{Loudon} However,
there are several experiments which challenge the conventional
view. Several x-ray absorption\cite{Garcia,Garcia1,8a,Bridges} and
neutron diffraction studies\cite{Daoud} revealed pictures that do
not match with the static mixture of Mn$^{3+}$ and Mn$^{4+}$ ions.
One suggests that all Mn ions have the same valence and
result into the Zener-polaron state.\cite{Ferrari,Zheng} However,
very recent experiment\cite{8a} observed the presence of two types
of Mn sites with different local geometric structures. One of the
types of Mn sites is surrounded by a tetragonal-distorted oxygen
octahedron, whereas the other
 has a regular octahedral
environment. As a result a charge segregation state was deduced.
With the motivation of the experimental observations\cite{8a} we model
the presence of the two types of Mn sites by incorporating the
Falicov-Kimball (FK) model.\cite{Falicov} The FK model was
initially introduced as a statistical model for metal-insulator
transition.\cite{Falicov} Later it was also applied to valence
change transitions in intermetallic
compounds.\cite{Zlatic,Farkasovsky} Within the FK model the
presence of two types of Mn sites is mapped to a classical
variable which only accept two values (for instance, $1$ and $0$).
The energy difference of these sites is mapped into the interaction
strength of the FK model. Indeed, the sites surrounded by
tetragonal-distorted octahedron have induced the JT distortion. As a
consequence the energy levels of the distorted Mn sites are split.
The FK model can describe a charge ordered phase as well as a
charge segregated phase.\cite{Freericks} In particular, the model
can exhibit the checkerboard CO state in appropriate conditions.
The checkerboard CO state is truly a mixed-valence state. The
segregated state is a phase-separated mixture of two full uniform
configurations.\cite{FreericksFalicov,Lemberger,Watson,Lemanski}  In
such the way, at low temperature the FK model could establish
various phases with different charge configurations which may
correspond to the experimental observations.\cite{8a}

However, the FK model alone cannot describe the DE induced FM
state which was also observed in doped manganites. Therefore we
incorporate the FK model into the DE model in order to study both
the charge ordered phases and the ferromagnetism upon doping. The
combined model has previously been considered in the context of
order-disorder change of the A-site substitution.\cite{Tran} In
the previous study\cite{Tran} only the checkerboard CO and FM state are
considered in the limit of infinite value of the Hund coupling.
In this paper we study all possible ordered phases of
the combined model in whole range of doping and interaction. In order to
detect the phase transition we study the static charge and spin
response of system within the dynamical-mean field theory
(DMFT).\cite{Georges} The DMFT has been widely used for
investigating strongly correlated electron systems. Within the
DMFT the static charge and spin correlation function are
calculated explicitly. We find that the system exhibits a rich
phase diagram which includes various charge ordered phases coexisting with
ferromagnetism. In particular, the checkerboard
CO state or the segregated state can coexist with the FM state.
The combined model can also serve as a model for studying the
problem of order-disorder A-site substitution \cite{Tran} or the
problem of orbital ordering in doped
manganites.\cite{Laad,Ferrari1}

The present paper is organized as follows: In Sec. II we present
the combined FK and DE model and its DMFT solutions. In Sec. III
the charge and spin correlation functions are calculated
explicitly. The numerical results and discussions are presented in
Sec. IV. The final section  is conclusion and remark.

\section{Model and Dynamical mean-field theory}

The combined FK and DE model in our study is described by the
following Hamiltonian
\begin{eqnarray}
H &=&-t\sum_{<ij>,\sigma }c_{i\sigma }^{\dagger }c_{j\sigma }^{\null}-\mu
\sum_{i\sigma }c_{i\sigma }^{\dagger }c_{i\sigma }^{\null}-2J_{H}%
\sum_{i}S_{i}^{z}s_{i}^{z}  \nonumber \\
&&+U\sum_{i\sigma }n_{i\sigma }w_{i}+E_{w}\sum_{i}w_{i},  \label{hamil}
\end{eqnarray}%
where $c_{i\sigma }^{\dagger }$($c_{i\sigma }^{\null}$) is the
creation (annihilation) operator for an itinerant electron with
spin $\sigma $ at lattice site $i$. The first term in
Hamiltonian (\ref{hamil}) represents the hoping of itinerant electrons
between the nearest neighbor sites. $t$ is the hoping integral and is scaled
with the spatial dimension $d$ as $t=t^{\star
}/(2\sqrt{d})$.\cite{Vollhardt} In the following we will take
$t^{\star }=1$ as the unit of energy. $S_{i}^{z}$ is the $z$
component of localized spin at lattice site $i$. For simplicity,
it takes two values $-1,1$.  $s_{i}^{z}=
(c_{i\uparrow}^{\dagger}c_{i\uparrow}-c_{i\downarrow}^{\dagger}c_{i\downarrow})/2$
is the $z$ component of itinerant
electron spin. $w_{i}$ is a classical variable that assumes the value
$1$($0$) if site $i$ is surrounded by distorted (regular) octahedron. $U$
is the interaction strength and is mapped into the difference in
the level energy of these sites. The expectation value
$\rho_{w}=\sum_{i}\langle w_{i}\rangle /N$, ($N$ is the number of lattice
sites), corresponds to the concentration of distorted sites. The
chemical potential $\mu $ controls the doping $n_{e}=\sum_{i\sigma
}\langle n_{i\sigma }\rangle /N$, while $E_{w}$ controls the
fraction of distorted sites. The condition $n_{e}+\rho _{w}=1$ is
used to determine $E_{w}$ for each doping $n_{e}$.
The first
three terms in Hamiltonian (\ref{hamil}) describe a simplified
DE model which contains only the Ising-type interaction between
the itinerant and localized electron spins. The simplification
does not allow any spin-flip processes, which can be important at
low temperature where spin-wave excitations may govern the
thermodynamics of the system. However, in the DE processes
spins of itinerant electrons align ferromagnetically with the
localized spins, hence the Ising part of the Hund coupling plays a
dominant role. Within the DMFT  the simplified DE model is
equivalent to the DE model with classical localized spins in the
disordered paramagnetic phase.\cite{Furukawa} The transport
quantities calculated within the simplified model capture
essential features of the full DE model.\cite{Nham} The simplified
DE model has also previously been used in the study of doped
manganites. \cite{Letfulov} The last two terms in Hamiltonian
(\ref{hamil}) take into account the energy difference of Mn sites.
They together with the hoping term form the FK
model.\cite{Falicov} Several authors have also constructed the combined model
to study the properties of manganites in different contexts and
regimes.\cite{Tran,Laad,Ferrari1,Ramak} Ferrari {\em et al.} used the combined model
to study the metallic FM phase of the two orbital DE model.\cite{Ferrari1}
Recently, Ramakrishnan {\em et al.} basically used the combined model
to construct a two band model of localized polaronic and broad
band states.\cite{Ramak} They used the DMFT to calculate transport
quantities and explained the metal insulator transition and the colossal
magnetoresistance in doped manganites.\cite{Ramak}

We solve the combined model (\ref{hamil}) by the
DMFT.\cite{Georges} The DMFT is based on the infinite dimension
limit. In the infinite dimension limit the self-energy is pure
local and does not depend on momentum. The Green function of
itinerant electrons with spin $\sigma$ satisfies the Dyson
equation
\begin{eqnarray}
G_{\sigma}(\mathbf{k},i\omega_n)=\frac{1}{i\omega_n -\epsilon (\mathbf{k}%
)+\mu -\Sigma_{\sigma}(i\omega_n)},
\end{eqnarray}
where $\epsilon (\mathbf{k})=-2t\sum_{i=1,d}\cos(k_{i})$ is the
dispersion of free itinerant electrons on a hypercubic
lattice, and $\Sigma_{\sigma}(i\omega_n)$ is the self energy which
depends only on frequency. The self energy is determined by
solving an effective single-site problem. The action for this
effective problem is
\begin{eqnarray}
S_{{\small {\text{eff}}}}=-\int_{0}^{\beta} d\tau \int_{0}^{\beta}
d\tau'\sum_{\sigma}c_{\sigma}^{\dagger}(\tau)\mathcal{G}%
_{\sigma}^{-1}(\tau -\tau')c_{ \sigma}(\tau')
\nonumber \\
-\int_{0}^{\beta} d\tau \sum_{\sigma}[J_HS^z\sigma + \mu
-Uw]c_{\sigma}^{\dagger}(\tau)c_{\sigma}(\tau) +  \beta E_w w,
\label{seff}
\end{eqnarray}
where $\mathcal{G}_{\sigma}(\tau -\tau')$ is the Green
function of the effective medium. It plays as the bare Green
function of the effective problem. The local Green function also
satisfies the Dyson equation
\begin{eqnarray}
G_{\sigma}^{-1}(i\omega_n)=\mathcal{G}_{\sigma}^{-1}(i\omega_n)-\Sigma_{%
\sigma}(i\omega_n),  \label{g22}
\end{eqnarray}
where $\mathcal{G}_{\sigma}(i\omega_n)$ is the Fourier transform
of $\mathcal{G}_{\sigma}(\tau )$. The local Green function
$G_{\sigma}(i\omega_n)$ of the effective single-site problem is
solely determined by the partition function
\begin{eqnarray}
G_{\sigma}(i\omega_n)=\frac{\displaystyle \delta
\mathcal{\ln Z}_{{\small {\text{eff}}}}} {\displaystyle \delta
\mathcal{G}_{\sigma}^{-1}(i\omega_n )} ,  \label{g122}
\end{eqnarray}
where $Z_{{\small {\text{eff}}}}$ is the partition function of the
effective problem (\ref{seff}). The self-consistent condition of
the DMFT requires that the local Green function
$G_{\sigma}(i\omega_n)$ obtained within the effective problem must
coincide with the local Green function of the original lattice,
i.e.,
\begin{eqnarray}
G_{\sigma}(i\omega_n)=\frac{1}{N}\sum_{\mathbf{k}}G_{\sigma}(\mathbf{k}%
,i\omega_n) \hspace*{2.5cm}  \nonumber \\
=\int d\epsilon \rho(\epsilon)\frac{1}{i\omega_n -\epsilon +\mu
-\Sigma_{\sigma}(i\omega_n)} , \label{g11}
\end{eqnarray}
where $\rho(\epsilon)$ is the density of state (DOS) of
noninteracting itinerant electrons. In the infinite dimension
limit of hypercubic lattices it has the form
$\rho(\epsilon)=\exp(-\epsilon^2/(t^{\star})^2)/\sqrt{\pi}t^\star$.
Eqs. (\ref{g22}), (\ref{g122}) and (\ref{g11}) form the
self-consistent equations for determining the self-energy, and
hence, also the Green function of the original lattice. Within the
effective single-site problem, the partition function is
\begin{eqnarray}
\mathcal{Z}_{{\small {\text{eff}}}}=\text{Tr}\int
Dc_{\sigma}^{\dagger}Dc_{\sigma}e^{-S_{{\small {\text{eff}}}}},
\end{eqnarray}
where the trace is taken over $S^{z}$ and $w$. This partition
function can be calculated exactly. It is similar to solve
the FK model within the DMFT.\cite{Brandt} We obtain
\begin{eqnarray}
\mathcal{Z}_{{\small {\text{eff}}}} =
2 \sum_{\alpha=0,1} \sum_{s=\pm 1} && \!\!\!\!\!\!\!
 \exp \Big[ \displaystyle -\beta
E_w \alpha +  \nonumber \\
&& \sum_{n\sigma} \ln \frac{Z_{\sigma}(i\omega_n)+\sigma s
J_H - \alpha U}{i\omega_n} \Big],
\end{eqnarray}
where $Z_{\sigma}(i\omega_n)\equiv
\mathcal{G}^{-1}_{\sigma}(i\omega_n)$. Using Eq. (\ref{g122}) we
obtain the local Green function
\begin{eqnarray}
G_{\sigma}(i\omega_n)=\sum_{\alpha s}\frac{W_{\alpha
s}}{Z_{\sigma}(i\omega_n)+\sigma s J_H - \alpha U} , \label{g33}
\end{eqnarray}
where the weight factors $W_{\alpha s}$ are
\begin{equation}
W_{\alpha s}=\frac{2}{\mathcal{Z_{{\small
{\text{eff}}}}}} \exp\Big[-\beta E_w\alpha+\sum_{n\sigma}\ln
\frac{Z_{\sigma}(i\omega_n)+\sigma s J_H -\alpha
U}{i\omega_n}\Big]
\end{equation}
with $\alpha=0,1$ and $s=\pm 1$.
Note that the weight factors $W_{\alpha s}$ are not simply a
number. They are functionals of the local Green function. This is
an important feature of the DMFT that gives nontrivial
contributions to the response functions of the
system.\cite{Brandt}  In the paramagnetic phase
$Z_{\uparrow}(i\omega_n)=Z_{\downarrow}(i\omega_n)$, hence
$W_{\alpha s}=W_{\alpha,-s}$ that leads the Green function
(\ref{g33}) and the self energy are independent of spin indeces,
as expected. The value of $E_w$ is adjusted that the concentration
$\rho_w$ fulfills $n_{e}+\rho _{w}=1$ for each doping $n_e$. One can
show that
\begin{equation}
\rho_{w}=\sum_{s=\pm 1} W_{1s} .
\end{equation}
We use this equation to adjust the value of $E_w$.
So far, we have obtained closed system of equations for determining the
Green function of the system. The system of equations can be solved
numerically by iterations.\cite{Freericks}

\section{Instability of homogeneous paramagnetic phase}

In order to detect the charge and spin ordered states which are established
at low temperature we study the static charge  and spin
correlation function of itinerant electrons in disordered
paramagnetic phase. The signal of a phase transition is a
divergence of these correlation functions at a certain
momentum. The charge (c) and spin (s) correlation function are
defined as
\begin{equation}
\chi^{\text{c(s)}}(i,j) = \big\langle ( \delta n_{i\uparrow} \pm \delta
n_{i\downarrow} ) ( \delta n_{j\uparrow} \pm \delta n_{j\downarrow} ) %
\big\rangle ,  \label{csdef1}
\end{equation}
where $\delta n_{i\sigma} = n_{i\sigma} - \langle n_{i\sigma}
\rangle$. These correlation functions can be expressed as
\begin{equation}
\chi^{\text{c(s)}}(i,j) = \sum_{\sigma \sigma'} \chi_{\sigma
\sigma'}(i,j)\xi_{\sigma} \xi_{\sigma'} , \label{csdef}
\end{equation}
where $\chi_{\sigma\sigma^{\prime}}(i,j) = \big\langle \delta
n_{i\sigma} \delta n_{j\sigma^{\prime}} \big\rangle$, and
$\xi_{\sigma}=1$ for the charge correlation function and
$\xi_{\sigma}=\sigma$ for the spin correlation function. In order
to calculate the static correlation functions, one has to
introduce external fields $h_{i\sigma}$ which couple to
$n_{i\sigma}$ into the Hamiltonian. The correlation functions
$\chi_{\sigma\sigma^{\prime}}(i,j)$ are obtained by differentiating
the Green function with respect to the external fields and then taking
the zero limit of the fields,\cite{Brandt} i.e.,
\begin{eqnarray}
\chi_{\sigma\sigma^{\prime}}(i,j) = - T^2 \sum_n \frac{d
G_{ii,\sigma}(i\omega_n)}{d h_{j\sigma^{\prime}}} \bigg
|_{\{h\}=0} . \label{correfc}
\end{eqnarray}
Following the standard technique,\cite{Brandt,Freericks} one can
express the charge and spin correlation function in the terms of
the charge and spin susceptibility
$\chi^{\text{c(s)}}(\mathbf{q},i\omega_n)$
\begin{equation}
\chi^{\text{c(s)}}(\mathbf{q}) = - T^2 \sum_n
\chi^{\text{c(s)}}(\mathbf{q}, i\omega_n) ,  \label{corr}
\end{equation}
where $\chi^{\text{c(s)}}(\mathbf{q})$ is the static charge (spin)
correlation function in momentum space. From the definition of
charge and spin correlation function (\ref{csdef}) and
relation~(\ref{correfc}), we obtain
\begin{eqnarray}
\chi^{\text{c(s)}}(\mathbf{q},i\omega_n) = 2 \chi_{0}(\mathbf{q},i\omega_n)
+ \mbox{\hspace{3.4cm}}  \nonumber \\
\chi_{0}(\mathbf{q},i\omega_n) \frac{1}{2} \sum_{\nu \sigma
\sigma'} \frac{d
\Sigma_{\sigma}(i\omega_n) }{d G_{\sigma'}(i\omega_\nu)} \xi_{\sigma} \xi_{\sigma'} \chi^{%
\text{c(s)}}(\mathbf{q},i\omega_\nu) ,  \label{chi}
\end{eqnarray}
where $\chi_{0}(\mathbf{q},i\omega_n)=\sum_{\bf k} G_{\sigma}({\bf
k + q},i\omega_n) G_{\sigma}({\bf k},i\omega_n)$ is the bare
particle-hole susceptibility. Here we have used the fact that in
the paramagnetic phase
\begin{eqnarray}
d \Sigma_{\uparrow}(i\omega_n) / d
G_{\uparrow}(i\omega_n)=d \Sigma_{\downarrow}(i\omega_n) / d
G_{\downarrow}(i\omega_n),
\end{eqnarray}
\begin{eqnarray}
d \Sigma_{\uparrow}(i\omega_n) / d
G_{\downarrow}(i\omega_n)=d \Sigma_{\downarrow}(i\omega_n) / d
G_{\uparrow}(i\omega_n).
\end{eqnarray}
From Eqs.~(\ref{g22}) and (\ref{g33}),
we obtain the self-energy $
\Sigma_\sigma(i\omega_n)$ as a functional of $G_{\sigma}(i\omega_n)$ and
$W_{\alpha s}$, hence its full derivatives in Eq.~(\ref{chi}) are expressed
through its partial derivatives. We obtain
\begin{eqnarray}
\frac{d \Sigma_{\sigma}(i\omega_n)}{d G_{\sigma'}(i\omega_\nu)}=
\delta_{n\nu} \delta_{\sigma \sigma'} \Big( \frac{\partial
\Sigma_{\sigma}(i\omega_n)}{\partial
G_{\sigma'}(i\omega_n)} \Big)_{W}  \nonumber \\
+\sum_{\alpha
s}\Big(\frac{\partial\Sigma_{\sigma}(i\omega_n)}{\partial
W_{\alpha s}}\Big)_{G,W_{\overline{\alpha s}}} \frac{\delta W_{\alpha s}}{%
\delta G_{\sigma'}(i\omega_n)},  \label{diffself}
\end{eqnarray}
where $W_{\overline{\alpha s}}$ means all weight factors $W$
except $W_{\alpha s}$. Substituting (\ref{diffself}) into
Eq.~(\ref{chi}) we arrive at
\begin{equation}
\chi^{\text{c(s)}}({\bf q},i\omega_n) =
 \frac{\displaystyle 2 + \frac{1}{2} \sum_{\alpha s \sigma}
\Big(  \frac{\partial \Sigma_{\sigma}(i\omega_n)}{\partial
W_{\alpha s}} \Big)_{G,W_{{\overline{\alpha s}}}} \xi_{\sigma}
\xi_{s} \gamma_{\alpha s}({\bf q})}{\displaystyle [\chi_{0}({\bf
q},i\omega_n)]^{-1} -\frac{1}{2} \sum_{\sigma} \Big(\frac{\partial
\Sigma_{\sigma}(i\omega_n)}{\partial G_{\sigma}} \Big)_{W} },
\label{chi1}
\end{equation}
where the matrix elements of $\widehat{\gamma}(\mathbf{q})$ are
\begin{equation}
\gamma_{\alpha s}(\mathbf{q}) = \sum_{\nu \sigma'}
\Big(\frac{\delta W_{\alpha s}}{\delta
G_{\sigma'}(i\omega_\nu)}\Big) \xi_{s} \xi_{\sigma'}
\chi^{\text{c(s)}}(\mathbf{q},i\omega_\nu) . \label{gamma}
\end{equation}
The functional derivative of $W_{\alpha s}$ in the above equation
can explicitly be expressed through the derivatives of $W_{\alpha
s}$ with respect to $Z_{\sigma}(i\omega_n)$ and the partial
derivatives of the self energy
$\Sigma_{\sigma}(i\omega_n)$.\cite{Brandt}
Substituting~(\ref{chi1}) into~(\ref{gamma}) and performing some
algebraic calculations we obtain the following matrix equation
\begin{equation}
\widehat{Q}(\mathbf{q}) \widehat{\gamma}(\mathbf{q}) =
\widehat{P}(\mathbf{q}) ,  \label{t}
\end{equation}
where the matrixes $\widehat{Q}(\mathbf{q})$,
$\widehat{P}(\mathbf{q})$ have the following elements
\begin{widetext}
\begin{eqnarray}
Q_{\alpha s, \alpha' s'}({\bf q}) &=& \delta_{\alpha s, \alpha'
s'} +
\nonumber \\
&& \sum_{n \sigma \sigma'} \bigg\{R_{\alpha s,\sigma}(i\omega_n)
S_{\sigma',\alpha' s'}(i\omega_n)
\Big(\frac{1}{2}-\delta_{\sigma\sigma'}\Big) - \frac{1}{2} \frac{
R_{\alpha s,\sigma}(i\omega_n) S_{\sigma',\alpha' s'}(i\omega_n)
\eta_n({\bf q}) G(i\omega_n) }{1 - G^{2}(i\omega_n) \Big(
 \frac{\displaystyle \partial \Sigma(i\omega_n)}{\displaystyle
 \partial G(i\omega_n) }
\Big)_{W} +  \eta_n({\bf q}) G(i\omega_n) }\bigg\} ,
\label{q} \\
P_{\alpha s}({\bf q}) &=& 2\sum_{n\sigma} \frac{ R_{\alpha
s,\sigma}(i\omega_n) \Big( G^{2}(i\omega_n) \Big(
 \frac{\displaystyle \partial \Sigma(i\omega_n)}
 {\displaystyle \partial G(i\omega_n) }
\Big)_{W} - 1 \Big)}{ 1 - G^{2}(i\omega_n) \Big(
 \frac{\displaystyle \partial \Sigma(i\omega_n)}
 {\displaystyle \partial G(i\omega_n) }
\Big)_{W} +  \eta_n({\bf q}) G(i\omega_n)
  } .
\label{p}
\end{eqnarray}
Here we have introduced the following notations
\begin{eqnarray*}
R_{\alpha s,\sigma}(i\omega_n)&=& \frac{\partial W_{\alpha s}}{
\partial Z_{\sigma}(i\omega_n)} \xi_{s} \xi_{\sigma} , \\
S_{\sigma,\alpha s}(i\omega_n)&=& \frac{\partial
\Sigma_{\sigma}(i\omega_n)}{\partial W_{\alpha s}} \xi_{\sigma}
\xi_{s} ,
\end{eqnarray*}
and $\eta_{n}(\mathbf{q})=-G^{-1}(i\omega_n)+
G(i\omega_n)\chi_{0}^{-1} (\mathbf{q},i\omega_n)$. In
Eqs.~(\ref{q})-(\ref{p}) the spin indices  of the Green function
and the self energy are omitted since they are in the paramagnetic
phase. The derivatives in Eqs.~(\ref{q})-(\ref{p}) can be
calculated explicitly. Straightforward calculations give
\begin{eqnarray}
\frac{\partial \Sigma_{\sigma}(i\omega_n)}{\partial W_{\alpha s}}
&=& \frac{1}{(Z(i\omega_n)+\sigma s J_{H}-\alpha U )
A_{\Sigma}(i\omega_n)}
,\label{diffSigma_W} \\
\frac{\partial \Sigma(i\omega_n)}{\partial G(i\omega_n)} &=& -
\frac{A_{G}(i\omega_n)}{A_{\Sigma}(i\omega_n)},
\label{diffSigma_G} \\
\frac{\partial W_{\alpha s}}{\partial Z_{\sigma}(i\omega_n)}&=&
\frac{W_{\alpha s}}{Z(i\omega_n)+\sigma s J_H -\alpha U}-W_{\alpha
s}G(i\omega_n) , \label{diffW_Z}
\end{eqnarray}
where
\begin{eqnarray*}
A_{\Sigma}(i\omega_n)&=& \sum_{\alpha s}
\frac{W_{\alpha s}}{(Z(i\omega_n)+\sigma s J_H - \alpha U)^{2}}, \\
A_{G}(i\omega_n)&=&
  -\sum_{\alpha s}W_{\alpha s}
  \frac{G^{-1}(i\omega_n)}{(Z(i\omega_n)+\sigma s J_H -\alpha
  U)^{2}}[Z(i\omega_n)+\sigma s J_H -\alpha U - G^{-1}(i\omega_n)].
\end{eqnarray*}
In such the way, Eqs. (19)-(24) fully determine the susceptibilities, once the self-consistent
equations of the DMFT are solved. The charge or spin correlation function
diverges whenever $\widehat{\gamma}(\mathbf{q})$
diverges, which happens when the determinant of $\widehat{Q}(\mathbf{q})$
vanishes. The divergence indicates the
instability of
 the disordered paramagnetic state. The $\mathbf{q}$
dependence of the susceptibilities
comes entirely from $\eta_n(\mathbf{q})$, and hence from the bare susceptibility
$\chi_0(\mathbf{q},i\omega_n)$.
Within the DMFT\cite{Freericks,Georges} in the infinite dimension limit all of the
$\mathbf{q}$ dependence
of the bare
susceptibility can be summarized in a single parameter
$X=\sum^{d}_{\alpha=1}\cos q_{\alpha}/d$.
One can show that\cite{Freericks,Georges}
\begin{eqnarray*}
\chi_0(\mathbf{q},i\omega_n)=\chi_0(X,i\omega_n)=-\frac{1}{\sqrt{1-X^2}}\int_{-\infty}^{+\infty}
d\epsilon \frac{\rho(\epsilon)}
{i\omega_n+\mu-\Sigma(i\omega_n)-\epsilon}  F_{\infty}\left(\frac
{i\omega_n+\mu-\Sigma(i\omega_n)-X\epsilon}{\sqrt{1-X^2}}\right),
\end{eqnarray*}
where $F_{\infty}(z)=\int d\epsilon\rho(\epsilon)/(z-\epsilon)$ is
used to denote the Hilbert transform. Now the instability of the
disordered paramagnetic phase happens whenever the determinant of
$\widehat{Q}(X)$ vanishes at a certain value $X$. In particular,
$X=-1$ corresponds to the checkerboard zone-boundary point
$\mathbf{q}=(\pi,\pi,...,\pi)$ and the corresponding instability
leads either to the charge checkerboard phase or to the
antiferromagnetic phase at low temperature. $X=1$ corresponds to
the uniform zone center point $\mathbf{q}=0$ and the corresponding
instability leads to a charge segregation phase or FM phase at low
temperature.
\end{widetext}

\section{Numerical results}

First, we consider the magnetic instability.
In this case we calculate the spin correlation function
as a function of $X$ and temperature $T$. A divergence of the spin
correlation function indicates a
magnetic instability.
For most values of $J_H$, $U$ and $n_e$ the spin correlation function diverges
only at $X=1$.
\begin{figure}[b]
\centerline{
\psfig{figure=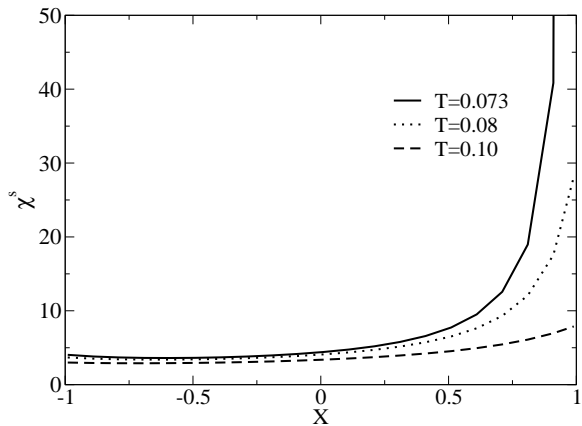,width=0.45\textwidth,angle=-90}
}
\caption{ The spin correlation function
as a function of $X$ at different temperatures ($n_e=0.5$, $J_H=2$, $U=0.5$).}
\label{h1b}
\end{figure}
In Fig. 1 we plot the typical behavior of the spin correlation function.
The divergence of the spin correlation function at $X=1$ indicates the
FM stability. This means the FM state is established at low
temperature. However for small values of $J_H$ and $n_e$ closed to
1, the spin correlation function also diverges at
$X=-1$.\cite{Letfulov3,Yunokia} This divergence indicates the
stability of the antiferromagnetic phase at low temperature. In
this paper we only consider the FM phase induced by the DE
mechanism and its coexistence with CO phases. Thus in the rest of
paper we consider the FM stability only.
\begin{figure}[b]
\centerline{ \psfig{figure=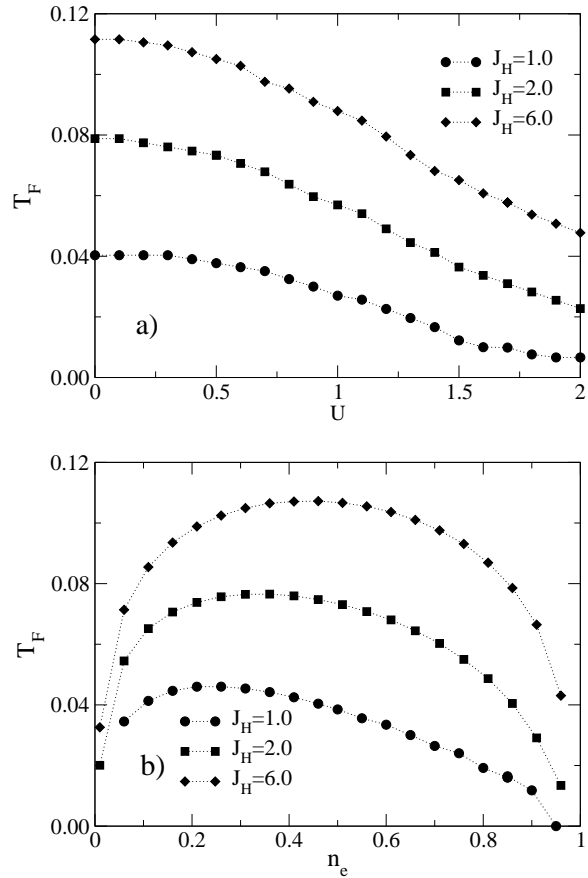,width=0.45\textwidth,angle=0}
} \caption{The FM transition temperature $T_F$ as a function of U
[(a)$n_e=0.5$] and as a function of $n_e$ [(b)U=0.5] for various
values of $J_H$.} \label{h2b}
\end{figure}
In Fig. 2 we present the FM transition temperature $T_F$ as a
function of $U$ and $n_e$ for various values of $J_H$. $T_F$ is
determined from the vanishing condition of the determinant of
$\widehat{Q}(X)$ at $X=1$. Fig. 2(a) shows that the critical
temperature $T_F$ decreases as increasing $U$ and increases as
increasing $J_H$. One expects in the limit $J_H\rightarrow \infty
$ the FM transition temperature reaches its maximum value for
fixed $U$. For fixed $J_H$ the FM transition temperature is
maximum if there is no JT distortion (i.e. $U=0$). The JT
distortion splits the energy level of Mn ions, and this leads to
suppress the FM transition temperature. This reduction of the FM
transition temperature due to distortion is gradually significant
already at intermediate values of $U$. In Figure 2(b) we also
present the FM transition temperature as a function of doping
$n_e$. In the limit $J_H\rightarrow \infty $, the FM  transition
temperature is maximum at half doping $n_e=0.5$.\cite{Tran}
However, for finite $J_H$ its maximum shifts away from the half
doping, to lower doping region.

Next, we consider the charge ordering instability.
In this case, we study the divergence of
the charge correlation function in the homogeneous paramagnetic phase.
\begin{figure}[t]
\centerline{ \psfig{figure=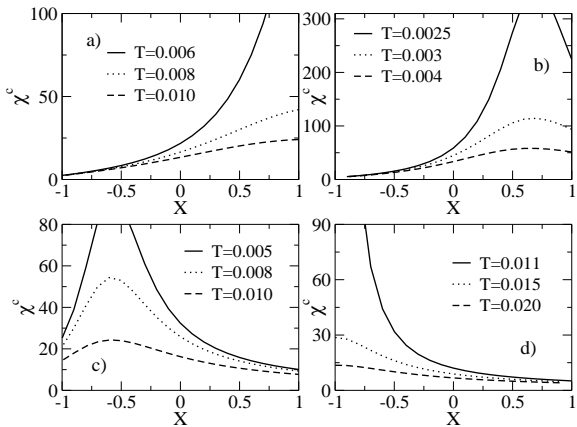,width=0.45\textwidth,angle=-90}
} \caption{The charge correlation function as a function of $X$ at
different temperatures. (a) $J_H=0.1$, (b) $J_H=0.2$, (c)
$J_H=0.4$, (d) $J_H=0.5$. In all figures $U=1.0$, $n_e=0.6$.}
\label{h3b}
\end{figure}
In Fig. 3 we plot the typical behaviors of the charge correlation
function. They show that the charge correlation function may
diverge at $X=1$, $X=-1$ or at an intermediate value $-1<X<1$. The
divergence at $X=-1$ indicates the checkerboard charge ordered
state established at low temperature, while the divergence at
$X=1$ indicates the segregated state established at low
temperature.  The divergence at an intermediate value of $X$
indicates the charge ordered phase being incommensurate at low
temperature. The charge ordering critical temperature $T_c$ is
determined  from the vanishing condition of the determinant of
$\widehat{Q}(X)$. However, one notices that for fixed values of
$J_H$, $U$ and $n_e$ the determinant of $\widehat{Q}(X)$ may
vanish at different $X$ and temperature $T$. Hence, we obtain the
critical temperature $T_c(X)$ as a function of $X$. However, this
does not indicate the stability of many charge ordered phases at
low temperature. For certain values of $J_H$, $U$ and $n_e$ there
is only one charge ordering transition which happens at the
maximum temperature $T_c$ among $T_c(X)$. Below this temperature
$T_c$, although the charge correlation function may still diverge
at other values of $X$, the divergence does not indicate a charge
ordering, because the assumption of disordered phase is not valid
anymore. Therefore the charge ordering critical temperature is
determined not only from the vanishing condition of the
determinant of $\widehat{Q}(X)$, but also from the maximum
condition\cite{Freericks}
\begin{eqnarray*}
T_c =\displaystyle\max_{\{X\}} T_c(X)
\end{eqnarray*}
\begin{figure}[t]
\centerline{
\psfig{figure=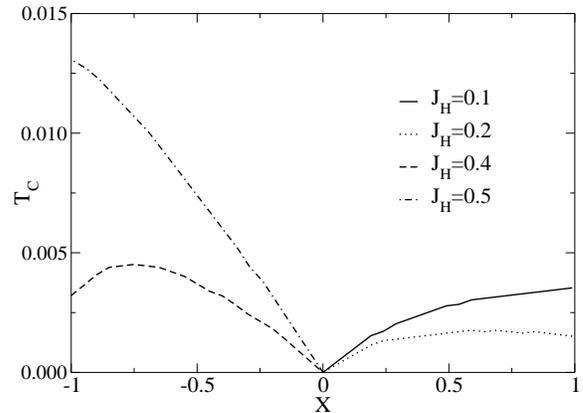,width=0.45\textwidth,angle=-90}
}
\caption{The dependence of $T_c$ on $X$ in case of $U=1.0$,
$n_e=0.6$ with various values of $J_H$.}
\end{figure}
\begin{figure}[b]
\centerline{
\psfig{figure=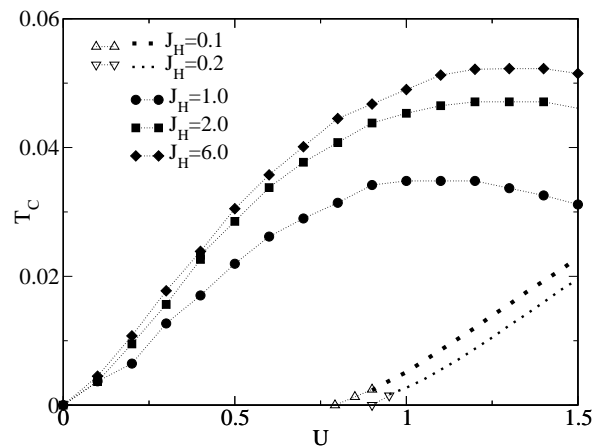,width=0.45\textwidth,angle=-90}
}
\caption{The critical temperature of charge ordered phase transition as a function
of $U$ ($n_e=0.5$). The filled (opened) symbols and the dotted lines correspond to the
checkerboard (incommensurate) charge ordered phase
and segregated phase, respectively.}
\label{h4b}
\end{figure}
The corresponding value of $X$ at which $T_c(X)$ is maximum determines
the charge arrangement of the low temperature phase. In
Fig. 4 we plot function $T_c(X)$ for different values of $J_H$.
It shows that for fixed $J_H$,
$U$, $n_e$ we always find a unique maximum $T_c$. For small values of $J_H$ we obtain the
segregated state at low temperature. For large values of $J_H$ the checkerboard ordered
phase is observed. In an intermediate regime we also find an incommensurate charge
ordering phase transition. In Fig. 5 we plot the charge ordering critical temperature
 as a function of $U$ for various values of $J_H$. It shows that the
checkerboard ordered phase is established for large values of
$J_H$. As increasing $U$, the critical temperature first
increases, reaches a maximum value, and then decreases. The
behavior of the critical temperature is similar to the one
obtained in the FK model.\cite{Brandt} For small values of $J_H$
the checkerboard CO phase disappears and the segregated phase or
incommensurate CO phase may be established depending on the value
of $U$, as shown in Fig. 5. Figure. 5 also shows that the
segregated phase is established at large values of $U$, while the
incommensurate phase is established at smaller values of $U$. One
notices that the segregated phase detected from the divergence of
the charge correlation function constitutes a continuous phase
transition. However, the phase transition is indeed first
order.\cite{Freericks,FreericksCh} It can be shown by considering
the free energy and using a Maxwell construction at low
temperature. \cite{Freericks,FreericksCh} Usually, the critical
temperature of the first order phase transition is higher than the
one obtained from the divergence of the charge correlation
function.\cite{Freericks,FreericksCh}

\begin{figure}[t]
\centerline{
\psfig{figure=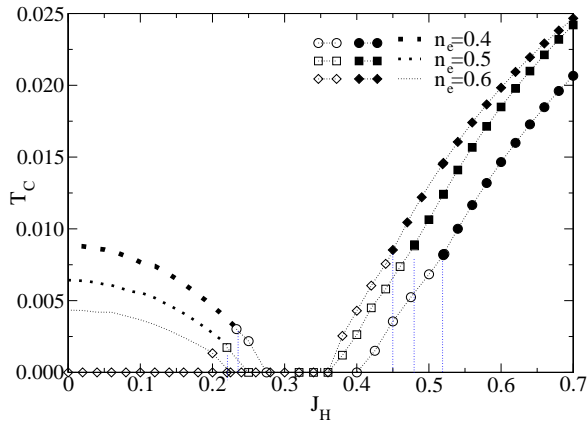,width=0.45\textwidth,angle=-90}
}
\caption{The charge ordering critical temperature $T_c$ as a function
of $J_H$ ($U=1.0$).
The filled (opened) symbols and the dotted lines correspond to the checkerboard (incommensurate)
charge ordered phase and the segregated phase, respectively.}
\label{h5b}
\end{figure}
\begin{figure}[t]
\centerline{
\psfig{figure=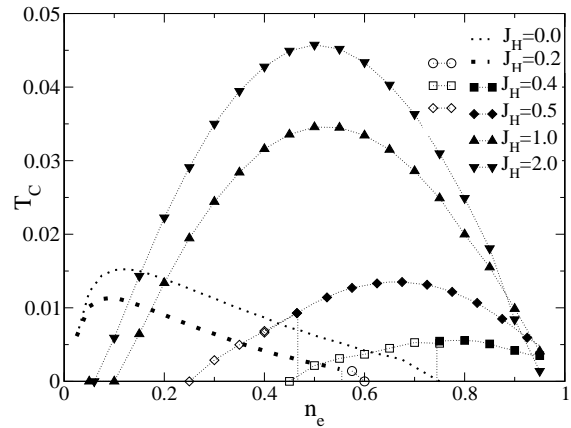,width=0.45\textwidth,angle=-90}
}
\caption{The charge ordering critical temperature $T_c$ as a function
of $n_e$ ($U=1.0$).
The filled (opened) symbols and the dotted lines correspond to the checkerboard (incommensurate)
charge ordered phase and the segregated phase, respectively.}
\label{h6b}
\end{figure}

In Fig. 6 we plot the critical temperature $T_c$ as a function
of $J_H$ for various doping $n_e$. It
shows that the checkerboard charge ordered phase is established at
large values of $J_H$ and disappears at small $J_H$. Its critical
temperature increases as increasing $J_H$. In contrast, the
segregated phase is established at small values of $J_H$ and its
critical temperature decreases as increasing $J_H$. For
intermediate values of $J_H$, both the checkerboard CO
and segregated phases disappear, and an incommensurate CO phase is
established. In Fig. 7 we plot the critical temperature $T_c$ as
a function of doping $n_e$ for
various values of $J_H$. For large values of $J_H$ the
checkerboard CO phase is established, and its critical
temperature reaches maximum at half doping $n_e=0.5$. For
intermediate values of $J_H$, the CO phase is
established at large doping $n_e>0.5$, and disappears at smaller
doping. Instead of the checkerboard CO phase, an
incommensurate phase appears. For small values of $J_H$ only the
segregated phase appears at small doping. At large doping the
segregated phase also disappears.

\begin{figure}[t]
\centerline{ \psfig{figure=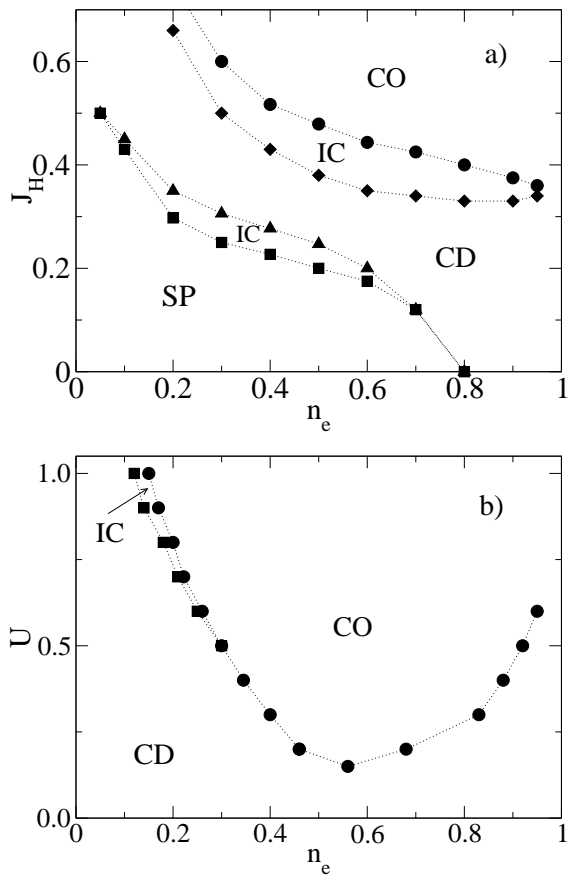,width=0.45\textwidth} }
\caption{Charge ordering phase diagrams: (a) $U=1$; (b) $J_H=1$.
The shorthand CO denotes the checkerboard charge ordered phase;
IC, the incommensurate phase; SP, the segregated phase, and CD,
the charge disordered phase.} \label{h10b}
\end{figure}

We summarize the above results with the phase diagram in Fig. 8
which plots the regions of stability for different charge ordered
phases. The stability is determined by the symmetry label $X$ of
the initial ordered phase as the temperature is lowered to the
first instability at $T_c$. Furthermore, we assume that the
symmetry label $X$ of the ordered phase does not change as the
temperature lowered from $T_c$ to zero. Actually, the phase
diagram is an approximation of the zero temperature phase
diagram.\cite{Freericks} The phase boundaries may change as one
reduces the temperature from $T_c$ to zero since the behaviors of
incommensurate phases at low temperature are not able to be
studied within the present approach. Moreover, the phase
boundaries may also change if there are first-order phase
transitions which may happen with the segregation phases. The
phase diagram shows that the incommensurate phases is stabilized
in buffer zones between the disordered and checkerboard CO phase
or between the disordered and segregated phase. The segregated
phase exists only for small values of $J_H$. For large values of
$J_H$ the checkerboard CO phase is stabilized. So far, we have
obtained different charge ordered phases depending on the value of
$J_H$, $U$ and doping $n_e$. On the other hand, the system always
exhibits the FM stability. Although the FM transition may happen
first, in the FM phase the charge density still remains
homogeneous and the charge correlation function is still normal as
in the homogeneous paramagnetic phase. Hence one can use the
charge instability signal in the high temperature homogeneous
phase as the signal of charge ordering even if the system is
already in the FM phase. Indeed, the charge ordering critical
temperature  detected from the charge order parameter in the CO-FM
state coincides with the one calculated from the instability of
the charge correlation function in the homogeneous paramagnetic
phase.\cite{Tran} In such the way, the FM state coexists with
different charge ordered phases for appropriate doping and
distortion. One notices that in the previous
studies\cite{Yunokia,Yunoki,Hotta,Yunokib} a rich phase diagram
which includes spin, charge and orbital ordered phases was also
obtained. There is also a coexistence of the FM phase with
checkerboard CO phase due to the Jahn-Teller
phonons.\cite{Yunokia,Yunoki,Hotta,Yunokib} In particular, the
ferromagnetic CO phase is stabilized for large Jahn-Teller
coupling and infinite Hund coupling. The conditions are in an
agreement with the phase diagram in Fig. 8. However, in the
previous studies\cite{Yunokia,Yunoki,Hotta,Yunokib} only a phase
separation between different magnetic phases was considered. A
separation of charge ordered phases has not been addressed. In the
present paper, a charge segregated phase coexisting with the FM
phase is obtained theoretically. The phase is a phase separated
mixture of two types of Mn sites with different local geometric
structures. The regime of the phase separation is particularly
interesting, and possible consequences of its existence may be
relevant to the experimental observations.\cite{8a} However, the
segregated phase appears only for weak Hund coupling and strong
distortion. There is also a possible coexistence of the two
segregated phases, one of which is between the magnetic phases,
and the other is between the charge ordered phases. However, we
leave the problem for further study.

\section{Conclusion}

In the present paper we have considered the doping and distortion
effect on the double-exchange ferromagnetism. By employing the
DMFT we have exactly calculated the charge and spin correlation
function. A long range order is determined from the divergence
signal of the correlation functions. The obtained results show
that the system exhibits various phases which include the FM,
checkerboard CO, incommensurate CO and segregated phases. In
particular, the FM phase can coexist with the checkerboard CO
phase for large values of $J_H$ and with the segregated phase for
small values of $J_H$. The incommensurate phases appear in the
buffer zones between the regions of the charge ordered phases with
different symmetries. It is interesting to note that experiments
have observed both the CO-FM state\cite{Loudon} and charge
segregated phase.\cite{8a} The phase separation between different
charge ordered phases is a novel regime in manganites. By
including the distortion effect in manganites via the
Falicov-Kimball interaction we have simulated the charge
segregated phase. The phase diagrams were found to clearly
distinguish regions with robust CO-FM correlations and charge
phase separation. However, manganites are too complicated of a
system to be completely described by this simple model. In
particular, experiments have observed inhomogeneous regions with
different long-range orders.\cite{Loudon} The study of the
properties is beyond the capacity of the present method.

\vfill

\begin{acknowledgements}

The authors grateful to Professor Jim Freericks for providing his papers and
fruitful discussions.
They also thank Professor Holger Fehske for valuable discussions.
This work is supported by the
National Program of Basic Research on Natural Science, Project 410104.

\end{acknowledgements}

\end{document}